\documentclass[twocolumn,aps,prl,floatfix,showpacs,superscriptaddress]{revtex4}
\usepackage{graphicx,amsmath}
\begin{document}
\title{Pattern Selection and Super-patterns in the Bounded Confidence Model}
\author{E.~Ben-Naim}
\affiliation{Theoretical Division and Center for Nonlinear Studies,
Los Alamos National Laboratory, Los Alamos, New Mexico 87545 USA}
\author{A.~Scheel}
\affiliation{School of Mathematics, University of Minnesota, 
Minneapolis, MN 55455, USA}
\begin{abstract}
  We study pattern formation in the bounded confidence model of
  opinion dynamics. In this random process, opinion is quantified by a
  single variable. Two agents may interact and reach a fair
  compromise, but only if their difference of opinion falls below a
  fixed threshold.  Starting from a uniform distribution of opinions
  with compact support, a traveling wave forms and it propagates from
  the domain boundary into the unstable uniform state. Consequently,
  the system reaches a steady state with isolated clusters that are
  separated by distance larger than the interaction range. These
  clusters form a quasi-periodic pattern where the sizes of the
  clusters and the separations between them are nearly constant.  We
  obtain analytically the average separation between clusters
  $L$. Interestingly, there are also very small quasi-periodic
  modulations in the size of the clusters. The spatial periods of
  these modulations are a series of {\it integers} that follow from
  the continued fraction representation of the {\it irrational}
  average separation $L$.
\end{abstract}
\maketitle

The so-called ``bounded-confidence'' model
\cite{wdan,hk,bkr,eb,gw,flpr,jl} and variants thereof have been widely
used to model opinion dynamics \cite{cfl,kb,ss,sl,mpt,mm,mbv} and have
attracted a considerable amount of interest .  The bounded confidence
model is appealing because it captures the tendency for reaching
compromise through social interactions, while also taking into account
a certain degree of conviction.  Numerical studies show that political
parties emerge in the bounded confidence model as a result of a
pattern formation process \cite{wdan,bkr,eb}.  In this letter, we
obtain analytically the wavelength governing this process.

We focus on a version of the bounded confidence model in which
opinions are quantified as discrete variables $1\leq n \leq N$. In
each interaction, two agents with opinions $n_1$ and $n_2$ change
their initial opinions by adopting the average opinion
\hbox{$(n_1,n_2)\to (\tfrac{n_1+n_2}{2},\tfrac{n_1+n_2}{2})$}; such a
compromise occurs only when the opinion difference is smaller than
some fixed threshold $|n_1-n_2|\leq \sigma$.  We set the threshold
$\sigma=2$ and exclude interactions between agents whose opinion
difference equals one, $|n_1-n_2|=1$, to ensure that opinions remain
discrete variables. In this simplified version of the bounded
confidence model, opinions change according to \cite{eb}
\begin{eqnarray}
\label{process}
(n-1,n+1)\to (n,n).
\end{eqnarray}
Clearly, this process conserves population and opinion.

Let $P_n(t)$ be the probability density of agents with opinion $n$ at
time $t$.  This density obeys the rate equation 
\begin{eqnarray}
\label{master}
\frac{d P_n}{dt}=2P_{n-1}P_{n+1}-P_n(P_{n-2}+P_{n+2}).
\end{eqnarray}
In writing this equation, we implicitly take the infinite population
limit. It is simple to check that Eq.~\eqref{master} conserves
population, $\sum_n P_n$, and opinion, $\sum_n n\,P_n$.
     
The initial distribution of opinions is uniform with compact support,
\begin{equation}
\label{ic}
P_n(0)=
\begin{cases}
0 & n<1, \\ 
1 & 1\leq n \leq N, \\
0 & N<n.
\end{cases}   
\end{equation}
We view the parameter $N$ as the opinion ``spectrum,'' and also note
that $N$ is the only parameter in the model.  The evolution equation
\eqref{master} is invariant under the scaling transformation $P\to
\alpha P$ and $t\to t/\alpha$ and hence, we may set the uniform
initial density to unity. This choice allows us to compare systems
with different opinion spectrums.

The nature of the interaction \eqref{process}, also reflected by the
evolution equation \eqref{master}, implies that any probability
density that satisfies $P_{n-1}P_{n+1}=0$ for all $n$ is
stationary. Clearly, there are infinitely many such steady-state
solutions. Starting from the (unstable) initial condition \eqref{ic},
the {\it deterministic} rate equation \eqref{master} evolves the
system toward one of those (stable) steady-state solutions \cite{bkr}.

\begin{figure}[t]
\includegraphics[width=.475\textwidth]{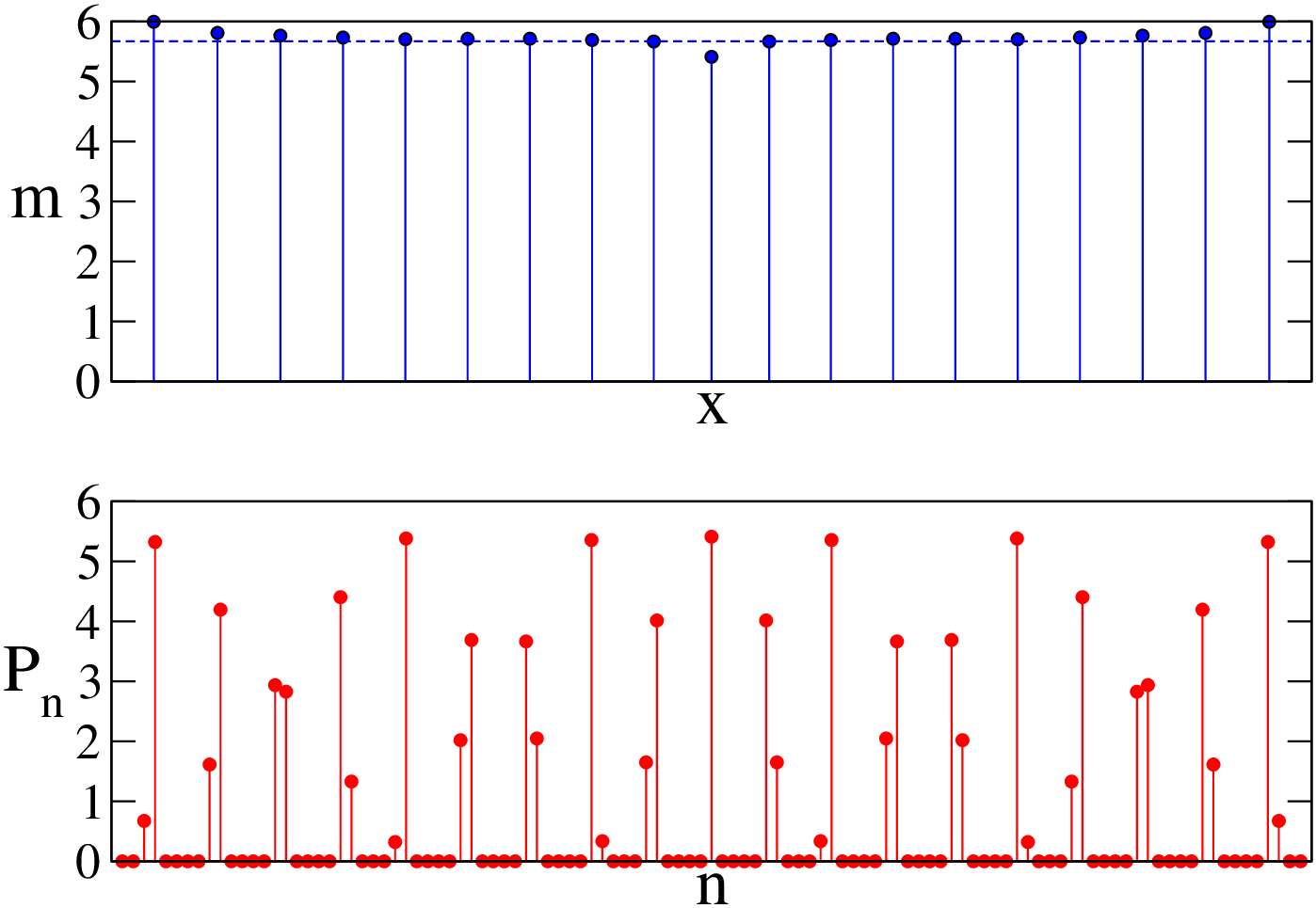}
\caption{[bottom] The probability density $P_n(\infty)$ versus $n$.
[top] The cluster mass $m$ versus position $x$. . Shown are results of
integration of \eqref{master}-\eqref{ic} with $N=109$. The dashed line
corresponds to the theoretical value $L=5.671820$}
\label{fig-frozen}
\end{figure}

In the final state (i.e., in the limit \hbox{$t\to\infty$}), the system
reaches a steady state where $P_{n-1}(\infty)P_{n+1}(\infty)=0$ for
all $n$.  In this state, there are multiple opinion ``clusters'' with
each cluster localized to two neighboring lattice sites (see figure
\ref{fig-frozen}). These clusters are noninteracting because the
separation between them exceeds the interaction range.

To quantify the size and opinion of each cluster, we compute for every
pair of occupied lattice sites the mass
\hbox{$m=P_{n}(\infty)+P_{n+1}(\infty)$} and the non-integer position
\hbox{$x=[nP_n(\infty)+(n+1)P_{n+1}(\infty)]/m$}.  Let $m_i$ be the
mass of the $i$th cluster and $x_i$ be the position of the $i$th
cluster. Conservation of population and opinion sets the sum rules
$\sum_i m_i=N$ and \hbox{$\sum_i m_ix_i=N(N+1)/2$}.

Unlike the probability density, the cluster mass forms a
quasi-periodic pattern (see figure \ref{fig-frozen}) as clusters of
nearly-identical mass are separated by a nearly-identical
distance. This pattern can be characterized by the average separation
$L$ between clusters.
\begin{equation}
\label{Ldef}
L=\lim_{N\to\infty} \langle m\rangle.
\end{equation}
With this definition, the average number of clusters scales as $N/L$ in
the limit $N\to\infty$. Previous numerical studies reported the value
$L\approx 5.67$ \cite{eb}. In this letter, we use theoretical methods
to analyze the evolution of the probability density $P_n$ and
analytically obtain $L$ as the wavelength that governs the underlying
pattern formation process.

The uniform initial state \eqref{ic} is unstable with respect to
perturbations that propagate from either boundary into the unstable
uniform state \cite{raf,ak}. By substituting the small periodic
disturbance
\begin{eqnarray}
\label{pert}
P_n(t)-1 \propto \exp[i(kn-\omega t)]
\end{eqnarray}
into the evolution equation \eqref{master} we find the dispersion
relation between frequency $\omega$ and wavelength $k$,
\begin{eqnarray}
\label{disp}
\omega=2i(2\cos k-\cos 2k-1).
\end{eqnarray}
Because the quantity $-i\omega$ is positive for $0<k<\pi/2$,
perturbations with wavenumber in that range initially grow
exponentially with time. The fastest growing mode, by ordinary linear
stability analysis, follows immediately from \eqref{disp}. The maximum
of $-i\omega$ in \eqref{disp} is set by \hbox{$d\omega/dk=0$} which
yields $k_{\rm linear}=\pi/3$ or alternatively \hbox{$L_{\rm
linear}=2\pi/k_{\rm linear}$}, that is, $L_{\rm linear}=6$.

\begin{figure}[t]
\includegraphics[width=.475\textwidth]{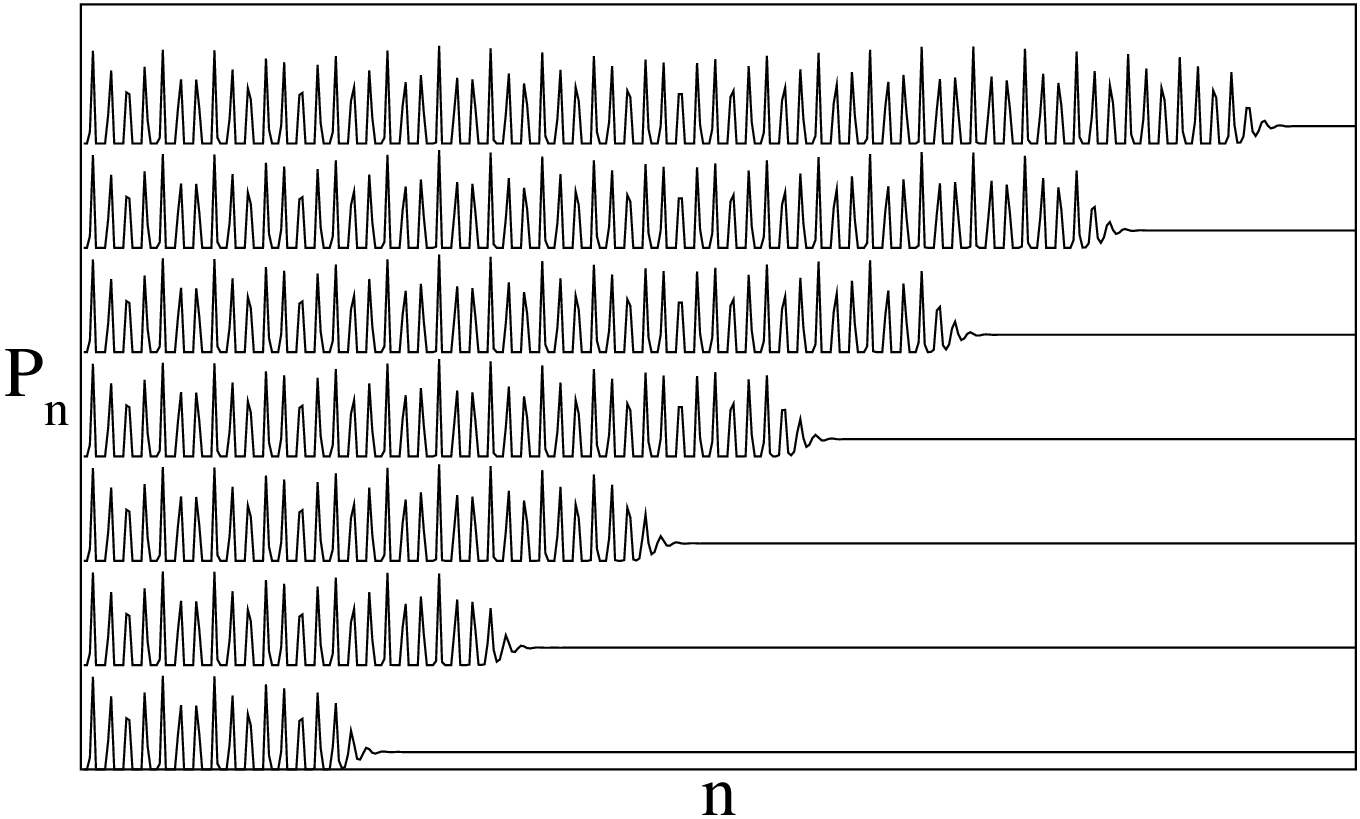}
\caption{Propagation from the stable into the unstable state. Shown is
$P_n(t)$ at equally spaced time intervals versus $n$. The curves
$P_n(t)$ are shifted vertically with the earliest time at the bottom
and the latest time on top.}
\label{fig-pn}
\end{figure}

The perturbations propagate from the stable state into the unstable
state at a constant velocity $v$ (see figure \ref{fig-pn}).  A saddle
point analysis shows that the propagation velocity $v$ obeys (for a
comprehensive review see \cite{vvs} and also \cite{kns,ev,gms,hs})
\begin{equation}
\label{velocity}
v=\frac{d\omega}{dk}=\frac{{\rm Im}[w]}{{\rm Im}[k]}.
\end{equation}
The solution to this equation is the complex wavenumber $k_*\equiv
k_{\rm front}+i\lambda$ with $k_{\rm front}=1.183032$ \cite{eb}.  The
constant $\lambda=0.467227$ characterizes the exponential decay
of these periodic perturbations \hbox{$P_n-1\sim e^{-\lambda
(n-vt)}\,e^{ik_{\rm front} (n-vt)}$}.  The wavelength of perturbations
at the leading edge of the front is $L_{\rm front}=2\pi/k_{\rm front}$ or 
explicitly,
\begin{equation}
\label{Lfront}
L_{\rm front}=5.311086.
\end{equation}
Moreover, the propagation velocity $v={\rm Im}[w_*]/{\rm Im}[k_*]$
where $w_*\equiv w(k_*)$ is
\begin{equation}
\label{v}
v=3.807397.
\end{equation}
Our numerical results confirm that in the leading edge of the
propagating front, the wavelength of the periodic deviations from the
uniform state is indeed given by \eqref{Lfront}.

\begin{figure}[t]
\includegraphics[width=.475\textwidth]{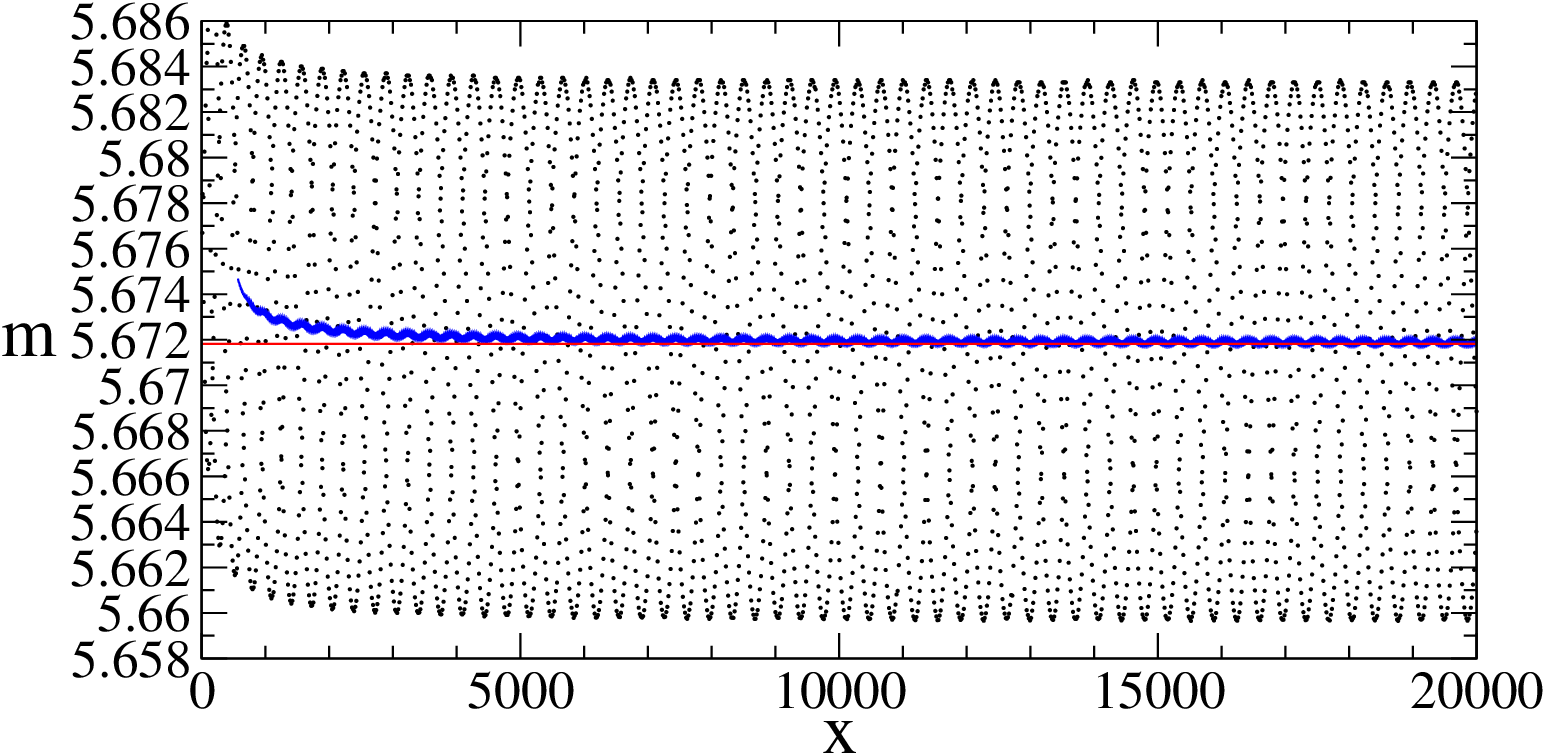}
\caption{The cluster mass $m$ versus position $x$.  The red line
corresponds to the theoretical value \eqref{Lpattern} and the blue
line, to a $100$-point running average.}
\label{fig-mx}
\end{figure}

Far behind the traveling wave, that is, in the wake of the wave, the
system reaches a steady state with $P_{n-1}P_{n+1}=0$ for all $n\ll
v\,t$. In this region, clusters are fully-developed. Interestingly,
far behind the propagating front, the pattern that forms, and which
ultimately controls the spacing between clusters, has a {\em larger}
wavelength due to a Doppler-like effect. The frequency of oscillations
in the front $w_*-k_*v$, measured in the co-moving frame, translates
to a zero frequency in the rest frame and hence, to stationary
patterns, precisely for the wavenumber \cite{dl,vvs,gms}
\begin{equation}
\label{kpattern}
k=k_*-\frac{w_*}{v}.
\end{equation}
We note that unlike the Doppler effect, the normalized shift in
wavenumber \hbox{$(k_*-k)/k_*$} does not equal a ratio of two
velocities as it is complex. The resulting wavenumber \eqref{kpattern}
is $k=1.107789$ and the corresponding wavelength $L=2\pi/k$ is
\begin{equation}
\label{Lpattern}
L=5.6718200283.
\end{equation}
Hence, out of the entire range of possible wavelengths corresponding
to linearly unstable perturbations, \hbox{$0<L<6$}, the wavenumber
\eqref{Lpattern} is ``selected'' by the dynamics of
Eq.~\eqref{master}.  We also note the inequality \hbox{$L_{\rm
front}<L<L_{\rm linear}$}.  Our numerical results give
excellent confirmation of the theoretical prediction (see
Fig.~\ref{fig-mx}): the numerically-measured wavelength
\hbox{$L=5.67185$} is within $10^{-5}$ of \eqref{Lpattern}.

To efficiently perform the computation, we integrated the equations
using a lattice with fixed size $\mathcal{N}$ that is moving at the
same speed as the traveling wave. Numerical integration in this {\it
co-moving} reference frame is feasible because far ahead of the
traveling wave $P_n=1$ and far behind it, the system settles into a
steady-state.  The lattice is shifted by one lattice site, $n\to n+1$
whenever the deviation from the uniform state at the extreme lattice
site, far ahead of the traveling front, exceeds an infinitesimal
threshold, $|P_\mathcal{N}-1|>\epsilon$. We used a Runge-Kutta (4,5)
integration method with adaptive step size below $10^{-3}$ on a
lattice of size $\mathcal{N}=2,000$ and the threshold
$\epsilon=10^{-250}$, well above the smallest machine precision
$10^{-308}$. To resolve the leading edge at this precision, we
integrated the equations for $Q_n=P_n-1$ where the leading edge decays
to the constant state $Q_n=0$.  Whenever the lattice is shifted by one
site, time and the probability density at the site $n=200$, well
behind the leading edge, were recorded.  This approach allows us to
integrate the equation to times $t\approx 3\times 10^6$ and
effectively, study systems with very large opinion spectrums $N\approx
10^6$.

\begin{figure}[t]
\includegraphics[width=.475\textwidth]{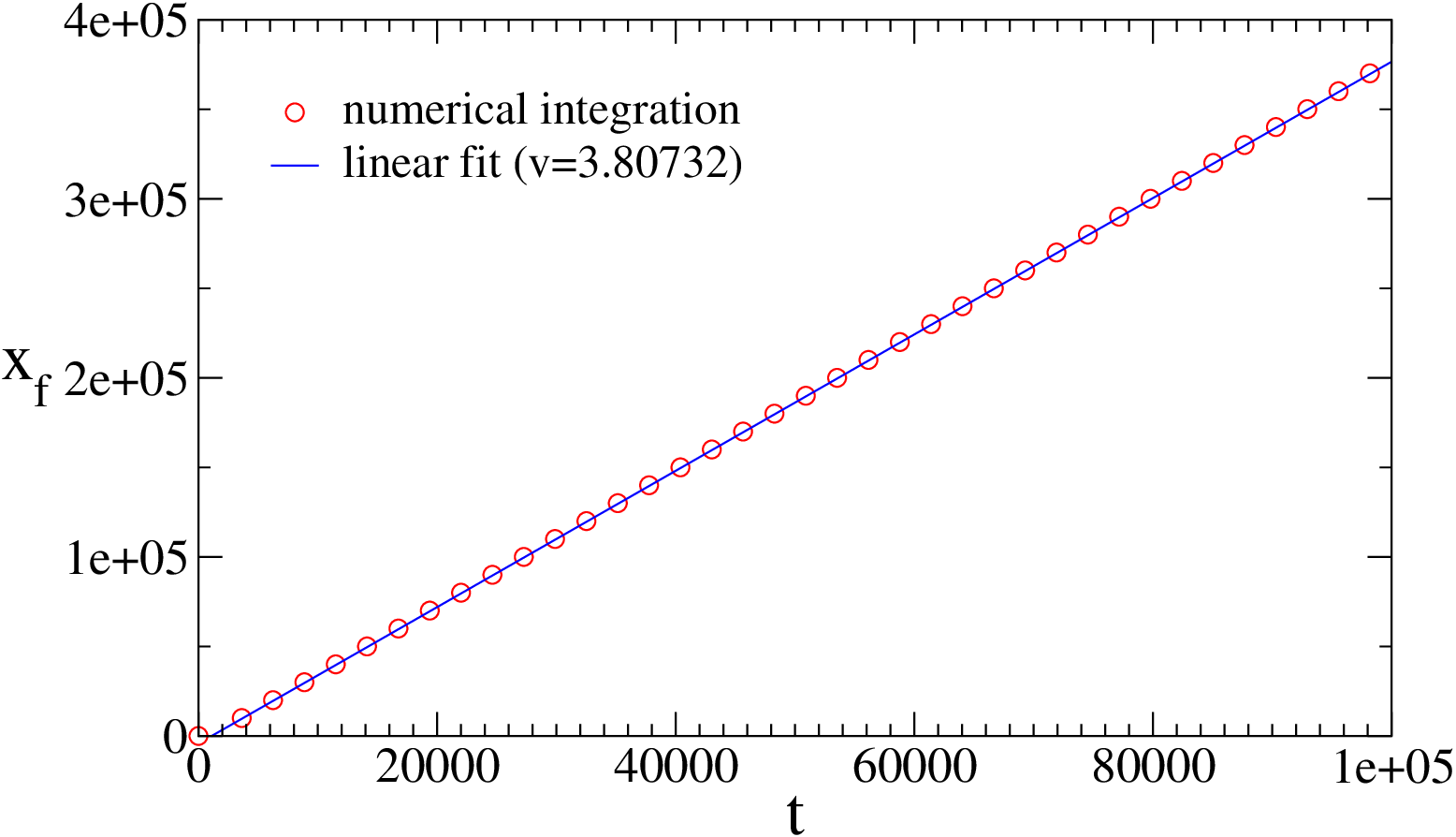}
\caption{The front location $x_f$ versus time $t$.}
\label{fig-xt}
\end{figure}

The number of shifts at time $t$ directly measures the front position
$x_f$. As shown in figure \ref{fig-xt}, The numerically-measured
propagation velocity $v=3.80732$ is in excellent agreement with the
theoretical prediction \eqref{v}. It is remarkable that our
computation, for which the ratio between system size and wavelength is
moderate, $\mathcal{N}/L\approx 350$, yields such high-precision
measurements of $L$ and $v$.  In general, a cutoff error $\epsilon$ in
the propagating front results in logarithmic correction $\delta \sim
(\ln \epsilon)^{-2}$ in the propagation velocity \cite{bd}. Moreover,
the cutoff decays exponentially with system size, $\epsilon \sim
\exp(-\lambda \mathcal{N})$. By combining these two scaling laws, we
find the algebraic relationship between system size $\mathcal{N}$ and
correction $\delta$ to the velocity, $\delta\sim (\lambda
\mathcal{N})^{-2}$.  The velocity correction we observe, $\delta
\approx 10^{-5}$ for $\mathcal{N}\approx 10^3$, is consistent with
this scaling law.

The irrational wavelength $L$ in \eqref{Lpattern} is not commensurate
with the unit lattice spacing. As a result, the patterns are not
strictly periodic but rather, they are quasi-periodic (see figure 
\ref{fig-mx}). Interestingly, we observed small but striking
``super-patterns'' induced by near-resonances between $L$ and the
lattice spacing. The integer periods of these super-patterns are found
from a continuous-fraction expansion of the wavelength
\begin{eqnarray}
\label{continued}
L=5+\frac{1}{1+\frac{1}{2+\frac{1}{21+\frac{1}{4+\ldots}}}}
               \equiv 6,\frac{17}{3},\frac{363}{64},\frac{1469}{259},\cdots.
\end{eqnarray}
Indeed, $3$ clusters can be accommodated within the integer period
$17$. Figure \ref{fig-pn} for the probability density $P_n$ shows
that a large peak in the quantity $P_n$ is usually followed by two
smaller ones. Hence, the system exhibits quasi-periodic behavior with
integer period $17$.

\begin{figure}[t]
\includegraphics[width=.475\textwidth]{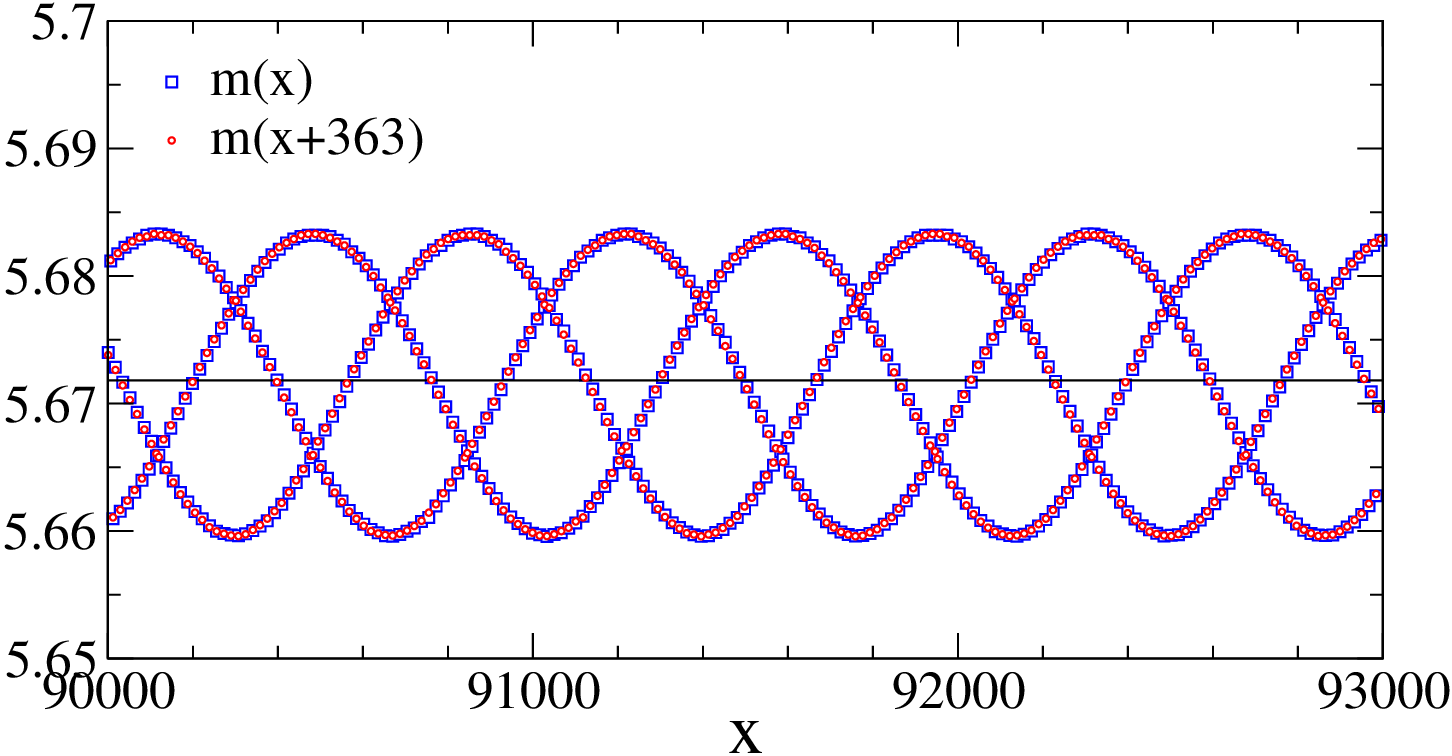}
\caption{The cluster mass $m$ versus position $x$ (large square) and
versus shifted position $x+363$ (small circle).  The line corresponds
to the theoretical value \eqref{Lpattern}.}
\label{fig-mx-pattern}
\end{figure}

Furthermore, there is also quasi-periodic arrangement of clusters with
integer period $363$.  Figure \ref{fig-mx-pattern} shows that the
cluster mass varies in the range \hbox{$L-\Delta < m < L+\Delta$}. The
variation in cluster mass is very small \hbox{$\Delta /m \approx
2\times 10^{-3}$}. As a function of position, these small variations
in cluster mass repeat with integer period $363$. According to the
continued fraction \eqref{continued}, $64$ clusters form this
pattern. This intriguing behavior was overlooked in previous studies
that used much smaller values of the opinion spectrum $N$
\cite{bkr,eb}.

Interestingly, the simple evolution equation \eqref{master} leads to a
hierarchical patterns, governed by a series of integer periods.  The
primary pattern, as shown in figure \ref{fig-frozen} consists of a
nearly-periodic arrangement of clusters with nearly-identical
separation. On the first hierarchical level, $M_1$ clusters are
arranged in a nearly periodic super-pattern with period $L_1$. On the
second hierarchical level, $M_2$ patterns form a more intricate
super-pattern with the integer period $L_2$. The fractions $M_1/L_1$,
$M_2/L_2$, and so on are rational approximations of the wavelength
$L$ that follow from its continued fraction
representation \eqref{continued}.

To examine the robustness of the above behavior, we also considered
the compromise process $(n-1,n+2)\to (n,n+1)$ where the interaction
range is three lattice sites.  In this case, the probability density
evolves according to the rate equation
\begin{eqnarray}
\label{master1}
\frac{d P_n}{dt}=P_{n-2}P_{n+1}+P_{n-1}P_{n+2}-P_n(P_{n-3}+P_{n+3}).
\end{eqnarray}
This equation conserves population and opinion. Starting from the
initial condition \eqref{ic}, the probability density evolves toward a
steady state where $P_n(\infty)P_{n+3}(\infty)=0$ for all $n$, and
therefore, opinion clusters are now localized to three consecutive
lattice sites.  By substituting \eqref{pert} into \eqref{master1}, the
dispersion relation is
\begin{eqnarray}
\label{dispersion1}
\omega=2i(\cos k+\cos 2k-\cos 3k-1).
\end{eqnarray}
By repeating the analysis leading to \eqref{Lpattern} we obtain the
average separation between clusters and propagation velocity 
\begin{equation}
\label{Lv}
L=8.5502770500\quad \text{and}\quad v=2.50631.
\end{equation}
Also, the wavelength of patterns nucleating at the front is $L_{\rm
  front}=8.02282$. Numerical integration of the evolution equation
gives $L=8.5503$ and $v=2.5063$, in excellent agreement with the
theoretical predictions.

\begin{figure}[t]
\includegraphics[width=.475\textwidth]{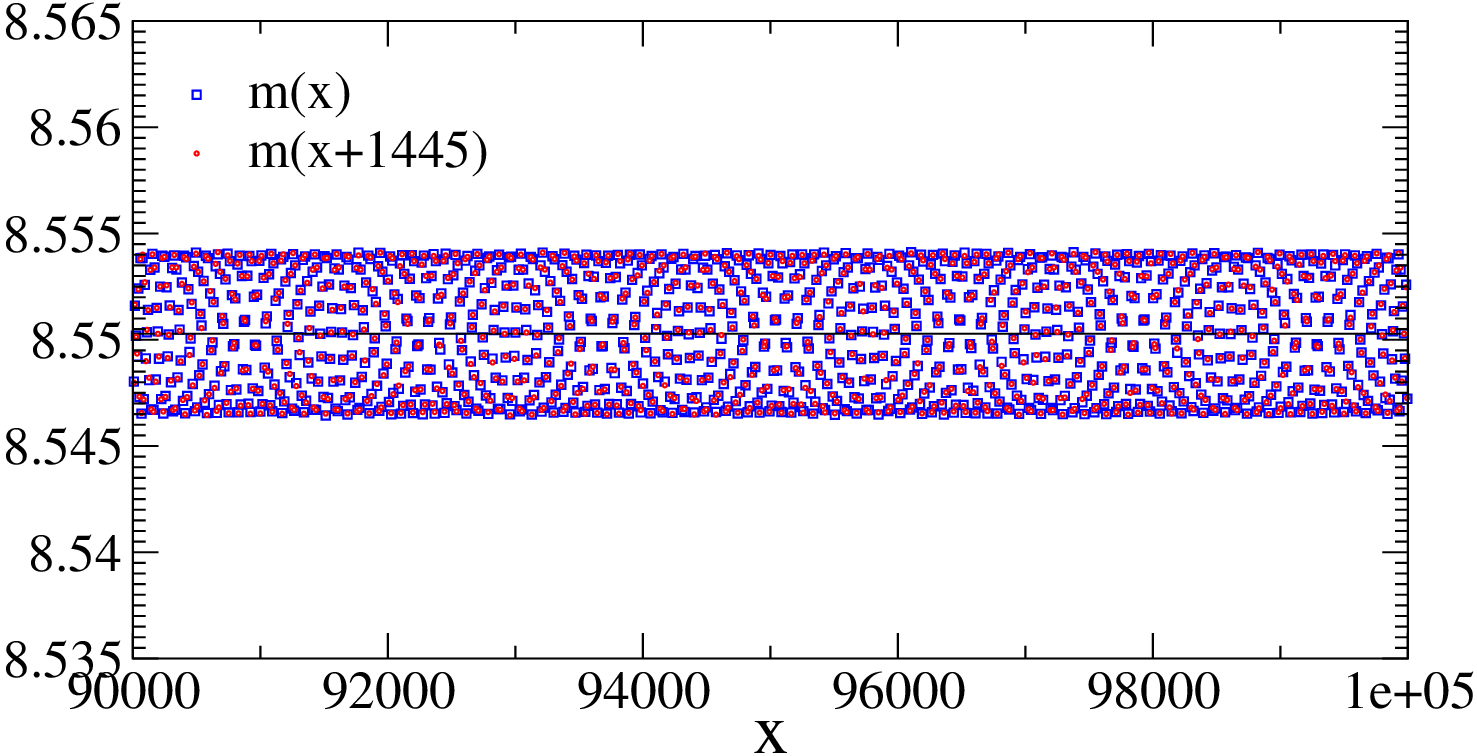}
\caption{The cluster mass $m$ versus position $x$ (large square) and
versus shifted position $x+1445$ (small circle) for the three-site
interaction model in \eqref{master1}.  The line corresponds to the
theoretical value \eqref{Lv}.}
\label{fig-mx-pattern-b}
\end{figure}

Figure \ref{fig-mx-pattern-b} demonstrates the emergence of
super-patterns.  In this case, continued fraction representation of
the wavelength \eqref{Lv} gives the rational approximations
\begin{eqnarray}
\label{continued-b}
L=8+\frac{1}{1+\frac{1}{1+\frac{1}{4+\frac{1}{2+\ldots}}}}
               \equiv9,\frac{17}{2},\frac{77}{9},\frac{171}{20},\frac{1445}{169}\cdots.
\end{eqnarray}
Figure \ref{fig-mx-pattern-b} shows that modulations in cluster mass
are periodic and well-characterized by the integer period $1445$, that
follows from the continued fraction \eqref{continued-b}. Accordingly,
the super-pattern consists of $169$ clusters. Furthermore, the
amplitude of the variations is very small, \hbox{$\Delta /m \approx
5\times 10^{-4}$}.

We also examined a linear interpolation between \eqref{master} and
\eqref{master1}, modeling a compromise process where second- or
third-neighbor interactions occur with relative weights $\tau$ and
$1-\tau$, respectively.  We found excellent agreement between the
theoretical predictions and the numerical results for the wavelength
$L$ and the velocity $v$ for all values of $\tau$.  As expected, the
wavelength decreases monotonically with the mixing parameter $\tau$,
which decreases the effective interaction range.  Surprisingly,
however, the velocity is not a monotonic function as it reaches a
minimum for $\tau \approx 0.88$. Thus, replacing a small fraction of
the second with third-neighbor interactions slows down the spreading
of the compromise process.

Moreover, we do not observe ``mode-locking''. In this phenomenon,
common in pattern forming systems that are exposed to an external
spatially-periodic forcing, there is pronounced locking of the
wavelength to that of the forcing \cite{db,wg}. In lattice systems,
one therefore expects the observed wavelength to be an integer, in
resonance with the unit lattice spacing, whenever the predicted
wavelength is close to an integer. In our case, the predicted
wavelength $L$ is integer ($6$ or $7$ or $8$) for particular values of
the mixing parameter $\tau$. We studied values of $\tau$ near those
resonances but did not observe any locking: the wavelength varies
smoothly and adheres to the predicted values. Further, we did not
observe subharmonic resonances as in \cite{s,bc,kp,mhhm}, where the
observed wavelength is an integer multiple of the predicted
wavelength.

We now briefly discuss the original bounded confidence model
introduced by Weisbuch et al \cite{wdan}.  In that model, opinion is
quantified by a continuous variable $0<x<N$ with $N$ the opinion
spectrum. Agents can interact and reach fair compromise but only if
their opinion difference falls below a fixed threshold, set to unity
without loss of generality \cite{bkr}
\begin{eqnarray}
\label{process1}
(x_1,x_2)\to \left(\tfrac{x_1+x_2}{2},\tfrac{x_1+x_2}{2}\right) \quad
\text{if} \quad |x_1-x_2|<1.
\end{eqnarray}
The probability density $P(x,t)$ of agents with opinion $x$ at time
$t$ obeys the evolution equation \cite{bkr}
\begin{eqnarray}
\label{master2}
{\partial\over \partial t}P(x,t)&=&\iint\limits_{|x_1-x_2|<1}
dx_1dx_2P(x_1,t)P(x_2,t)\nonumber\\
&\times&\left[\delta\left(x-{x_1+x_2\over 2}\right)-\delta(x-x_1)\right]
\end{eqnarray}
If the restriction of the integration range is ignored, this equation
describes inelastic collisions \cite{bk00,bk06,ska,rli}.  According to
the interaction \eqref{process1}, opinion clusters are now perfectly
localized (delta-functions) and in the steady-state these localized
clusters are separated by distance larger than unity.

Consider the uniform initial condition: $P(x,0)=0$ for $x<0$ or $x>N$
and $P(x,0)=1$ for $0\leq x\leq N$. This state is unstable with
respect to perturbations that propagate from the boundary into the
unstable uniform state. According to \eqref{master2}, a small periodic
disturbance $P(x,t)-1\propto \exp[i(kx-\omega t)]$ has the dispersion
relation
\begin{eqnarray}
\label{disp1}
\omega=2i\left[2\,\frac{\sin(k/2)}{k/2}-\frac{\sin k}{k}-1\right].
\end{eqnarray}
The fastest growing mode follows from $d\omega/dk=0$ which yields
$k_{\rm linear}=2.7906$ or alternatively \hbox{$L_{\rm
linear}=2.2515$}.  The solution to \eqref{velocity} is now $k_*\equiv
k_{\rm front}+i\lambda$ with \hbox{$k_{\rm front}=3.083750$}.  The
decay constant $\lambda=1.294620$ characterizes the exponential decay
far into the unstable state, $\phi(x)\sim \exp[-\lambda (x-vt)]$. The
wavelength of perturbations at the propagating front is \hbox{$L_{\rm
front}=2.037514$}.  The propagation velocity $v={\rm Im}[w_*]/{\rm
Im}[k_*]$ is \hbox{$v=0.794754$}.  Far behind the propagating front,
that is $x\ll vt$, localized clusters form, and these clusters are
separated by distance $L$. The corresponding wavenumber is
\hbox{$k=2.924255$} and the corresponding wavelength is
\begin{equation}
\label{Lcont}
L=2.1486444707.
\end{equation}
The wavelength estimated using numerical integration results for
relatively small values of $N$ \cite{bkr}, $L\approx 2.155$, is
reasonably close to the exact result \eqref{Lcont}.

In summary, we studied pattern formation in the bounded confidence
model of opinion dynamics. Our focus was the wavelength that governs
the mosaic of frozen clusters that develop, starting from a uniform
state.  We obtained analytically the two wavelengths that govern the
pattern formation process: the wavelength of perturbations at the
leading edge of the traveling wave front and the wavelength of the
resulting patterns in the wake of the wave.  We examined discrete and
continuous versions of the bounded confidence model. In the former
case, we verified the theoretical predictions using high-precision
numerical measurements of the pattern wavelength and propagation
velocity.

The wavelength of the patterns is irrational and since it is not
commensurate with the regular lattice, the pattern formation process
is hierarchical. Frozen clusters constitute the ``building bocks'' in
this hierarchy.  Integer number of clusters form quasi-periodic
structures and the period of these super-patterns is an integer, too.
Next, a larger number of clusters form a more intricate super-pattern
with a larger integer period.  The numbers of clusters and the periods
that characterize these super-patterns follow from continued fraction
representation of the irrational wavelength governing the pattern
formation process.

We observed that not all rational approximations of the wavelength
necessarily correspond to a super-pattern (see figure
\ref{fig-mx-pattern-b}). Further analysis is therefore needed to
understand why certain integer fractions are realized while others are
not, and more generally, to characterize the intricate structures of
the super-patterns.

\bigskip
  
We acknowledge financial support from the US Department of Energy
through grant DE-AC52-06NA25396 and from the US National Science
Foundation through grants DMS-0806614 and DMS-1311740.

\end{document}